# 基於晶格後量子密碼學同態加密

Abel C. H. Chen[*]

Information & Communications Security Laboratory,
Chunghwa Telecom Laboratories

## 摘要

　　隨著量子計算技術的日益成熟，具備抵抗量子計算攻擊的後量子密碼學方法也成為發展重點方向之一。有鑑於現行的同態加密方法包含 RSA、ElGamal、Paillier 等密碼學方法可能存在被量子計算攻破的風險，所以本研究設計基於晶格後量子密碼學同態加密，通過基於晶格密碼學方法建構抵抗量子計算攻擊的能力，同時又能提供同態加密計算應用。本研究提供數學理論證明和計算實例說明，以及探討基於晶格後量子密碼學同態密的安全性分析，後續可以提供給同態加密應用(例如：聯邦學習)開發者參考。

*關鍵詞：後量子密碼學、基於晶格密碼學、同態加密*

# Homomorphic Encryption Based on Lattice Post-Quantum Cryptography

**Abstract**

　　As quantum computing technology continues to advance, post-quantum cryptographic methods capable of resisting quantum attacks have emerged as a critical area of focus. Given the potential vulnerability of existing homomorphic encryption methods, such as RSA, ElGamal, and Paillier, to quantum computing attacks, this study proposes a lattice-based post-quantum homomorphic encryption scheme. The approach leverages lattice cryptography to build resilience against quantum threats while enabling practical homomorphic encryption applications. This research provides mathematical proofs and computational examples, alongside a security analysis of the lattice-based post-quantum homomorphic encryption scheme. The findings are intended to serve as a reference for developers of homomorphic encryption applications, such as federated learning systems.

---

[*] Corresponding author E-mail: chchen.scholar@gmail.com

# 1.前言

近幾年量子計算技術快速發展，並且隨著量子計算硬體技術成熟和錯誤率的降低，將可能搭配量子計算演算法來產生殺手級應用。例如，Google 在 2024 年 12 月於《Nature》期刊上發表論文"Quantum Error Correction Below the Surface Code Threshold"可以做到大幅降低量子計算錯誤率，並且實現指數級加速[1]。有鑑於此，美國國家標準暨技術研究院(National Institute of Standards and Technology, NIST)近幾年陸續訂定後量子密碼學標準[2]-[3]，並且在 2024 年 8 月發佈基於模晶格金鑰封裝機制標準(Module-Lattice-Based Key-Encapsulation Mechanism Standard, ML-KEM)[4]-[5]、基於模晶格數位簽章標準(Module-Lattice-Based Digital Signature Standard, ML-DSA)[6]-[7]、以及無狀態雜湊數位簽章標準(Stateless Hash-Based Digital Signature Standard, SLH-DSA)[8]-[9]。並且，美國國家標準暨技術研究院在 2024 年 12 月的" Transition to Post-Quantum Cryptography Standards"草案中呼籲大家在 2030 年之前開始汰換 RSA 密碼學和橢圓曲線密碼學等，遷移到後量子密碼學，以建構抵抗計算攻擊的能力，避免重要資料被竊取[10]。

有鑑於此，本研究在晶格密碼學的基礎上，設計基於晶格後量子密碼學同態加密，通過晶格密碼學的特性，避免被量子計算攻破。同時，本研究設計的基於晶格後量子密碼學同態加密可以提供加法同態加密，未來將可以被應用於聯邦學習等場域。

本論文總共包含四個章節。第 2 節主要基於晶格後量子密碼學和同態加密的基本構想。第 3 節介紹本研究設計的基於晶格後量子密碼學同態加密，並且提供數學理論證明和安全性分析。最後，第 4 節總結本研究貢獻。

# 2.文獻探討

在第 2.1 節將先介紹基於晶格後量子密碼學，解釋後量子密碼學的金鑰產製、加密、解密之原理，並且輔以計算實例說明。在第 2.2 節將定義和介紹同態加密方法，並以基於 RSA 的同態加密方法為例說明。

## 2.1 基於晶格後量子密碼學

本節將先說明基於晶格後量子密碼學的演算法設計及其原理，再舉計算實例展示。

## 2.1.1 演算法設計

本研究主要採用 NTRU 加密演算法[11]-[12]作為基於晶格後量子密碼學,以下說明 NTRU 加密演算法。

首先,將先產製整數 $p$ 和整數 $q$,並且最大公因數為 1 (即 $GCD(p, q) = 1$),以及多項式的階 $N$,然後產製 3 個多項式環,分別為 $R$、$R_p$、$R_q$,如公式(1)、公式(2)、公式(3)所示。其中,整數 $p$ 將影響明文編碼的值域,整數 $q$ 將影響密文編碼的值域,而多項式的階 $N$ 將影響可以放置多少個元素。並且,在公式(1)、公式(2)、公式(3)中的除號視為模數計算。以公式(1)為例,即 $R$ 為多項式 $\mathbb{Z}[x]$ 模多項式 $x^N - 1$ 的計算結果,$a_i$ 為 $x^i$ 的係數值。

$$R = \frac{\mathbb{Z}[x]}{(x^N - 1)} = \mathbb{Z}[x] (\mod x^N - 1) \equiv \sum_{i=0}^{N-1} a_i x^i. \tag{1}$$

$$R_p = \frac{\mathbb{Z}_p[x]}{(x^N - 1)} = \mathbb{Z}[x] (\mod p)(\mod x^N - 1). \tag{2}$$

$$R_q = \frac{\mathbb{Z}_q[x]}{(x^N - 1)} = \mathbb{Z}[x] (\mod q)(\mod x^N - 1). \tag{3}$$

在產製金鑰對的過程中,將先產製一個隨機多項式 $f(x)$ (如公式(4)所示),再基於多項式 $f(x)$ 產製多項式 $F_p(x)$ 和多項式 $F_q(x)$,需符合公式(5)和(6)定義,並且把 $\{f(x), F_p(x), F_q(x)\}$ 作為私鑰;其中,$f(x)$ 的編碼是 $a_{f,0}||a_{f,1}||\ldots||a_{f,N-1}$,$F_p(x)$ 的編碼是 $a_{Fp,0}||a_{Fp,1}||\ldots||a_{Fp,N-1}$,$F_q(x)$ 的編碼是 $a_{Fq,0}||a_{Fq,1}||\ldots||a_{Fq,N-1}$。為了產製公鑰,將先產製另一個隨機多項式 $g(x)$ (如公式(7)所示),並且與多項式 $F_q(x)$ 相乘後得到多項式作為公鑰 $h(x)$,$h(x)$ 的編碼是 $a_{h,0}||a_{h,1}||\ldots||a_{h,N-1}$,如公式(8)所示。

$$f(x)(\text{in } R) \equiv \sum_{i=0}^{N-1} a_{f,i} x^i. \tag{4}$$

$$F_p(x)(\text{in } R_p) = f(x)^{-1}(\text{in } R_p) \equiv \sum_{i=0}^{N-1} a_{Fp,i} x^i. \tag{5}$$

$$F_q(x)(\text{in } R_q) = f(x)^{-1}(\text{in } R_q) \equiv \sum_{i=0}^{N-1} a_{Fq,i} x^i. \tag{6}$$

$$g(x)(\text{in } R) \equiv \sum_{i=0}^{N-1} a_{g,i} x^i. \tag{7}$$

$$\begin{aligned} h(x)\big(\text{in } R_q\big) &= F_q(x)g(x)\big(\text{in } R_q\big), \\ &= \sum_{i=0}^{N-1} \sum_{j=0}^{N-1} a_{Fq,i} x^i a_{g,j} x^j \pmod{q}(\bmod\ x^N - 1), \\ &\equiv \sum_{i=0}^{N-1} a_{h,i} x^i. \end{aligned} \tag{8}$$

在加密的過程中，為了達到選擇明文攻擊下的不可區分性(INDistinguishability under Chosen Plaintext Attack, IND-CPA)或選擇密文攻擊下的不可區分性(INDistinguishability under Chosen Ciphertext Attack, IND-CCA)的安全性[13]，在每次加密時都會加入隨機數，讓同一公鑰對同一明文加密後，每次的密文可以不同。因此，首先將產製隨機多項式$r(x)$ (如公式(9)所示)，然後把明文解碼為多項式的係數值$\{a_{m,0}, a_{m,1}, …, a_{m,N-1}\}$，產製待加密訊息多項式$m(x)$ (如公式(10)所示)。透過公式(11)，可以運用公鑰$h(x)$對待加密訊息多項式$m(x)$進行加密，產製多項式$c(x)$作為密文，密文可以編碼為$a_{c,0}||a_{c,1}|| … ||a_{c,N-1}$。後續如果有需要儲存和傳送時，可以傳送多項式$c(x)$的係數值。其中，由於每次產製的隨機多項式$r(x)$不同，所以可以提升密文的安全性。

$$r(x)(\text{in } R) \equiv \sum_{i=0}^{N-1} a_{r,i} x^i. \tag{9}$$

$$m(x)\big(\text{in } R_p\big) \equiv \sum_{i=0}^{N-1} a_{m,i} x^i. \tag{10}$$

$$\begin{aligned} c(x)\big(\text{in } R_q\big) &= p \times h(x) \times r(x) + m(x) \pmod{q}(\bmod\ x^N - 1), \\ &= \left(p \sum_{i=0}^{N-1} \sum_{j=0}^{N-1} a_{h,i} x^i a_{r,j} x^j\right) + \left(\sum_{i=0}^{N-1} a_{m,i} x^i\right) \pmod{q}(\bmod\ x^N - 1), \\ &\equiv \sum_{i=0}^{N-1} a_{c,i} x^i. \end{aligned} \tag{11}$$

基於晶格後量子密碼學同態加密

...在解密的過程中，為了消除隨機多項式$r(x)$和保留待加密訊息多項式$m(x)$，所以通過公式(12)、公式(13)、公式(14)計算。其中，公式(12)主要將密文$c(x)$和私鑰$f(x)$相乘後模 $q$，然後公式(13)可以把公式(12)結果$t(x)$轉換到$R$環和$R_p$環的多項式$\tau(x)$，然後公式(14)可以把公式(13)結果$\tau(x)$和私鑰$F_p(x)$相乘後模 $p$，即可還原得到待加密訊息多項式$m(x)$。詳細數學理論證明如公式(12)、公式(13)、公式(14)所示。

$$\begin{aligned} t(x)\bigl(\text{in } R_q\bigr) &= c(x) \times f(x) \ (\text{mod } q)(\text{mod } x^N - 1), \\ &= \sum_{i=0}^{N-1}\sum_{j=0}^{N-1} a_{C,i} x^i a_{f,j} x^j \ (\text{mod } q)(\text{mod } x^N - 1), \\ &= ph(x)r(x)f(x) + m(x)f(x) \ (\text{mod } q)(\text{mod } x^N - 1), \\ &= pF_q(x)g(x)r(x)f(x) + m(x)f(x) \ (\text{mod } q)(\text{mod } x^N - 1), \\ &= pr(x)g(x) + m(x)f(x) \ (\text{mod } q)(\text{mod } x^N - 1), \\ &\equiv \sum_{i=0}^{N-1} a_{t,i} x^i. \end{aligned} \quad (12)$$

$$\tau(x)\bigl(\text{in } R_p\bigr) = t(x)(\text{mod } p) \equiv \sum_{i=0}^{N-1} a_{\tau,i} x^i. \quad (13)$$

$$\begin{aligned} p(x)\bigl(\text{in } R_p\bigr) &= \tau(x) \times F_p(x) \ (\text{mod } p)(\text{mod } x^N - 1), \\ &= \sum_{i=0}^{N-1}\sum_{j=0}^{N-1} a_{\tau,i} x^i a_{Fp,j} x^j \ (\text{mod } p)(\text{mod } x^N - 1), \\ &= pr(x)F_p(x) + m(x)f(x)F_p(x) \ (\text{mod } p)(\text{mod } x^N - 1), \\ &\equiv m(x) \equiv \sum_{i=0}^{N-1} a_{m,i} x^i. \end{aligned} \quad (14)$$

### 2.1.2 計算實例

本節提供 NTRU 加密演算法實例說明；其中，相關參數值包含如下：$N = 7$、$p = 3$、$q = 128$，值得注意的是這只是計算示意，在真實應用要採用更大的參數值來提升安全性和明文長度。

- 5 -

為了產製私鑰，根據公式(4)定義隨機產製多項式$f(x)$ (如公式(15)所示)，並且根據公式(15)定義隨機產製多項式$F_p(x)$和多項式$F_q(x)$，需符合公式(5)和(6)定義，如公式(16)和公式(17)所示。其中，$f(x)$的編碼是1||−1||1||0||0||−1||1，$F_p(x)$的編碼是0||2||0||0||1||0||1，$F_q(x)$的編碼是87||58||81||54||36||67||2，並且私鑰的編碼是$f(x)||F_p(x)||F_q(x)$。

$$f(x)(\text{in } R) \equiv x^6 - x^5 + x^2 - x + 1. \tag{15}$$

$$F_p(x)(\text{in } R_p) \equiv x^6 + x^4 + 2x. \tag{16}$$

$$\begin{aligned}&F_q(x)(\text{in } R_q)\\&\equiv 2x^6 + 67x^5 + 36x^4 + 54x^3 + 81x^2 + 58x + 87.\end{aligned} \tag{17}$$

為了產製公鑰，產製另一個隨機多項式$g(x)$ (如公式(18)所示)，並且與多項式$F_q(x)$相乘後得到多項式作為公鑰$h(x)$，$h(x)$的編碼是12||94||20||56||123||124||83，如公式(19)所示。

$$g(x)(\text{in } R) \equiv x^3 - x^2 + x - 1. \tag{18}$$

$$\begin{aligned}&h(x)(\text{in } R_q)\\&= (2x^6 + 67x^5 + 36x^4 + 54x^3 + 81x^2 + 58x + 87)\\&\quad (x^3 - x^2 + 2x - 1) \;(\text{mod } 128)(\text{mod } x^7 - 1)\\&\equiv 83x^6 + 124x^5 + 123x^4 + 56x^3 + 20x^2 + 94x + 12.\end{aligned} \tag{19}$$

假設有明文編碼為1||1||0||0||0||0||0，並且解碼為待加密訊息多項式$m(x)$ (如公式(20)所示)。在加密的過程中，產製隨機多項式$r(x)$ (如公式(21)所示)，以保證每次加密結果不同。通過公式(11)計算，可以得到密文$c(x)$，如公式(22)所示，並且密文$c(x)$可以編碼為98||18||58||119||126||82||13。

$$m(x)(\text{in } R_p) \equiv x + 1. \tag{20}$$

$$r(x)(\text{in } R) \equiv x^5 + 127x^4 + x^3 + 127. \tag{21}$$

$$c(x)\,(\text{in } R_q)$$
$$= 3 \times (83x^6 + 124x^5 + 123x^4 + 56x^3 + 20x^2 + 94x + 12)$$
$$\times (x^5 + 127x^4 + x^3 + 127) + (x + 1) \pmod{128}\pmod{x^7 - 1},$$
$$\equiv 13x^6 + 82x^5 + 126x^4 + 119x^3 + 58x^2 + 18x + 98. \tag{22}$$

在解密的過程中,首先根據公式(12)把密文$c(x)$乘上私鑰$f(x)$相乘後模$q$,如公式(23)所示。再通過公式(24)把多項式轉換到環,以及運用公式(14)把多項式$\tau(x)$乘上私鑰$F_p(x)$相乘後模$p$,即可還原得到待加密訊息多項式$m(x)$,如公式(25)所示。

$$t(x)\,(\text{in } R_q)$$
$$= (13x^6 + 82x^5 + 126x^4 + 119x^3 + 58x^2 + 18x + 98)$$
$$(x^6 - x^5 + x^2 - x + 1) \pmod{128}\pmod{x^7 - 1},$$
$$\equiv 9x^6 + 118x^5 + 6x^4 + 123x^3 + 3x^2 + 127. \tag{23}$$

$$\tau(x)\,(\text{in } R_p)$$
$$\equiv 9x^6 + 118x^5 + 6x^4 + 123x^3 + 3x^2 + 127 \pmod{128},$$
$$\equiv 9x^6 - 10x^5 + 6x^4 - 5x^3 + 3x^2 - 1 \pmod{128},$$
$$\equiv 2x^5 + x^3 + 2 \pmod{3}. \tag{24}$$

$$p(x)\,(\text{in } R_p) = (2x^5 + x^3 + 2)(x^6 + x^4 + 2x) \pmod{3}\pmod{x^7 - 1},$$
$$\equiv x + 1 \equiv m(x). \tag{25}$$

## 2.2 同態加密方法

本節將先定義和說明同態加密方法的演算法設計,再舉計算實例展示。由於同態加密可分為全同態加密(Fully Homomorphic Encryption, FHE)和半同態加密(Partial Homomorphic Encryption, PHT),而全同態加密可同時支援加法同態加密(Additive Homomorphic Encryption, AHE)和乘法同態加密(Multiplicatively

Homomorphic Encryption, MHE)，但半同態加密則只能支援加法同態加密或乘法同態加密之一[14]-[15]。為簡化說明，本節主要介紹基於 RSA 乘法同態加密方法。

## 2.2.1 定義和演算法設計

首先定義乘法同態加密方法，令加密函數為$e(a_i)$，解密函數為$d(c_i)$，分別如公式(26)和公式(27)所示。其中，$a_i$為明文，$c_i$為密文。在乘法同態加密方法中，加密函數和解密函數同時符合公式(28)和公式(29)的定義。表示密文相乘後的結果$\prod_{i=1}^{n} c_i$解密後可以得到明文相乘後的結果$\prod_{i=1}^{n} a_i$；如此可以做到在密文時做計算，再把密文計算後的結果做一次解密即可得到結果。

$$c_i = e(a_i). \tag{26}$$

$$a_i = d(c_i). \tag{27}$$

$$\prod_{i=1}^{n} c_i = \prod_{i=1}^{n} e(a_i). \tag{28}$$

$$\prod_{i=1}^{n} a_i = d\left(\prod_{i=1}^{n} c_i\right). \tag{29}$$

本節採用基於 RSA 乘法同態加密方法來說明。在 RSA 密碼學中，假設私鑰為$\zeta$，公鑰為$\xi$，階為$\kappa$。因此，可以通過公式(30)對明文$a_i$加密得到密文$c_i$，以及通過公式(31)對密文$c_i$解密得到明文$a_i$，並且符合$a_i^{\zeta\xi} \pmod{\kappa} \equiv a_i$特性。RSA 密碼學的加密函數和解密函數可以符合公式(28)和公式(29)的特性，$d(\prod_{i=1}^{n} c_i) \equiv \prod_{i=1}^{n} a_i$，如公式(32)和公式(33)所示。因此，RSA 密碼學可以用來建構基於 RSA 乘法同態加密方法。

$$e(a_i) \equiv a_i^{\xi} \pmod{\kappa} \equiv c_i. \tag{30}$$

$$d(c_i) \equiv c_i^{\zeta} \pmod{\kappa} \equiv a_i^{\zeta\xi} \pmod{\kappa} \equiv a_i. \tag{31}$$

$$\prod_{i=1}^{n} e(a_i) \equiv \prod_{i=1}^{n} c_i \pmod{\kappa} \equiv \prod_{i=1}^{n} a_i^{\xi} \pmod{\kappa} \equiv \left(\prod_{i=1}^{n} a_i\right)^{\xi} \pmod{\kappa}. \tag{32}$$

$$d\left(\prod_{i=1}^{n} c_i\right) \equiv \left(\prod_{i=1}^{n} c_i\right)^{\zeta} \pmod{\kappa} \equiv \left(\left(\prod_{i=1}^{n} a_i\right)^{\xi}\right)^{\zeta} \pmod{\kappa}$$
$$\equiv \left(\prod_{i=1}^{n} a_i\right)^{\xi\zeta} \pmod{\kappa} \equiv \prod_{i=1}^{n} a_i. \tag{33}$$

### 2.2.2 計算實例

本節提供基於 RSA 乘法同態加密方法實例說明；其中，假設產製的金鑰值包含如下：$\zeta = 59$、$\xi = 3$、$\kappa = 391$，值得注意的是這只是計算示意，在真實應用要採用更大的參數值來提升安全性。假設有兩個明文，分別為 $a_1 = 11$、$a_2 = 13$，並且明文加密後分別為密文 $c_1 = 158$、$c_2 = 242$，如公式(34)和公式(35)所示。通過公式(36)可以計算密文相乘結果為 309，再通過公式(37)解密可以得到明文相乘結果 143。

$$e(a_1) \equiv 11^3 \pmod{391} \equiv 158 = c_1. \tag{34}$$

$$e(a_2) \equiv 13^3 \pmod{391} \equiv 242 = c_2. \tag{35}$$

$$\prod_{i=1}^{2} c_i \pmod{\kappa} \equiv 158 \times 242 \pmod{391} \equiv 309. \tag{36}$$

$$d\left(\prod_{i=1}^{2} c_i\right) \equiv 309^{59} \pmod{391} \equiv 143 = 11 \times 13 = \prod_{i=1}^{2} a_i. \tag{37}$$

## 3. 本研究設計的基於晶格後量子密碼學同態加密

本研究設計的基於晶格後量子密碼學同態加密主要適用於加法同態加密，所以在第 3.1 節定義加法同態加密的特性。第 3.2 節描述基於晶格後量子密碼學同態加密的設計理念，並且提供數學理論證明，以及在第 3.3 提供計算實例。最後，第 3.4 節進行安全性討論。

## 3.1 加法同態加密定義

本節定義加法同態加密方法，令加密函數為$E(a_i)$，解密函數為$D(c_i)$，分別如公式(38)和公式(39)所示。其中，$a_i$為明文，$c_i$為密文。在加法同態加密方法中，加密函數和解密函數同時符合公式(40)和公式(41)的定義。表示密文加總後的結果$\prod_{i=1}^{n} c_i$解密後可以得到明文加總後的結果$\prod_{i=1}^{n} a_i$；如此可以做到在密文時做計算，再把密文計算後的結果做一次解密即可得到結果。

$$c_i = E(a_i). \tag{38}$$

$$a_i = D(c_i). \tag{39}$$

$$\sum_{i=1}^{n} c_i = \sum_{i=1}^{n} E(a_i). \tag{40}$$

$$\sum_{i=1}^{n} a_i = D\left(\sum_{i=1}^{n} c_i\right). \tag{41}$$

## 3.2 設計理念

本節將先闡述金鑰產製，再說明加密和解密計算，以及提供數學證明本方法在加法同態加密的可行性。

### 3.2.1 金鑰產製

本研究設計的基於晶格後量子密碼學同態加密，主要建構在 NTRU 加密演算法的基礎上，所以需先先產製整數 $p$ 和整數 $q$，並且最大公因數為 1 (即 $GCD(p, q) = 1$)，以及多項式的階 $N$，然後產製 3 個多項式環，分別為$R$、$R_p$、$R_q$，如公式(1)、公式(2)、公式(3)所示。之後，在 3 個多項式環的基礎上，產製私鑰$\{f(x), F_p(x), F_q(x)\}$，如公式(4)、公式(5)、公式(6)所示。最後，再根據私鑰$F_q(x)$和產製隨機多項式$g(x)$來產製公鑰$h(x)$，如公式(8)所示。

### 3.2.2 加密

本節中假設有 $n$ 筆資料分別可以解碼為在$R_p$環的多項式，如第 $s$ 筆資料解碼為$m_s(x)$，如公式(42)所示。加密過程中為達到 IND-CPA 或 IND-CCA 安全等級，為每一筆資料各別產製不同的隨機多項式，如第 $s$ 筆隨機多項式為$r_s(x)$，如公式

(43)所示。之後再以公式(44)產製密文，如第 $s$ 筆資料的密文為$c_s(x)$。在加法同態加密計算上，可對密文直接加總，所以 $n$ 筆資料的密文加總結果為$C(x)$，如公式(45)所示。

$$m_s(x)(\text{in } R_p) \equiv \sum_{i=0}^{N-1} a_{m_s,i} x^i, \text{ where } s \in \{1,2,\ldots,n\}. \tag{42}$$

$$r_s(x)(\text{in } R) \equiv \sum_{i=0}^{N-1} a_{r_s,i} x^i, \text{ where } s \in \{1,2,\ldots,n\}. \tag{43}$$

$$\begin{aligned}
E(m_s, r_s) &= c_s(x)\big(\text{in } R_q\big), \\
&= p \times h(x) \times r_s(x) + m_s(x) \pmod{q} \pmod{x^N - 1}, \\
&= \left( p \sum_{i=0}^{N-1} \sum_{j=0}^{N-1} a_{h,i} x^i a_{r_s,j} x^j \right) + \left( \sum_{i=0}^{N-1} a_{m_s,i} x^i \right) \pmod{q} \pmod{x^N - 1}, \\
&\equiv \sum_{i=0}^{N-1} a_{c_s,i} x^i.
\end{aligned} \tag{44}$$

$$\begin{aligned}
\sum_{s=1}^{n} E(m_s, r_s)\big(\text{in } R_q\big) &= \sum_{s=1}^{n} c_s(x)\big(\text{in } R_q\big) \\
&= \sum_{s=1}^{n} [p \times h(x) \times r_s(x) + m_s(x)] \pmod{q} \pmod{x^N - 1}, \\
&= \sum_{s=1}^{n} [ph(x)r_s(x)] + \sum_{s=1}^{n} [m_s(x)] \pmod{q} \pmod{x^N - 1}, \\
&\equiv \sum_{i=0}^{N-1} a_{C,i} x^i \equiv C(x)\big(\text{in } R_q\big).
\end{aligned} \tag{45}$$

### 3.2.3 解密

解密的計算上與公式(12)、公式(13)、公式(14)相同，主要乘上私鑰多項式$f(x)$，然後轉換到$R_p$環，再乘上私鑰多項式$F_p(x)$，如公式(46)所示。

$$D\left(\sum_{s=1}^{n} c_s(x)\right)(\text{in } R_p) = D\big(C(x)\big)(\text{in } R_p),$$
$$= [C(x) \times f(x)(\text{in } R_q)] \times F_p(x)(\text{in } R_p). \tag{46}$$

### 3.2.4 理論證明

本節提出公式(47)證明 $n$ 筆資料的密文加總結果 $\sum_{s=1}^{n} E(m_s, r_s)\,(\text{in } R_q)$ 解密後可以得到 $n$ 筆資料的明文加總結果 $\sum_{s=1}^{n} m_s(x)\,(\text{in } R_p)$。其中,雖然每筆密文中有隨機多項式為 $r_s(x)$ 的影響,來達到 IND-CPA 或 IND-CCA 安全等級,但由於有乘上整數 $p$,所以轉換到 $R_p$ 環時,隨機多項式為 $r_s(x)$ 影響將會被消除。因此,最終僅會留下明文加總結果。

$$\begin{aligned}
D\left(\sum_{i=1}^{n} c_i\right)(\text{in } R_p) &= D\big(C(x)\big)(\text{in } R_p), \\
&= [C(x) \times f(x)(\text{in } R_q)] \times F_p(x)(\text{in } R_p), \\
&= \left[\left(\sum_{s=1}^{n}[ph(x)r_s(x)] + \sum_{s=1}^{n}[m_s(x)]\right)f(x)(\text{in } R_q)\right] \\
&\quad \times F_p(x)(\text{in } R_p), \\
&= \left[\left(\sum_{s=1}^{n}[pF_q(x)g(x)r_s(x)f(x)] + \sum_{s=1}^{n}[m_s(x)f(x)]\right)(\text{in } R_q)\right] \\
&\quad \times F_p(x)(\text{in } R_p), \\
&= \left(\sum_{s=1}^{n}[pg(x)r_s(x)] + \sum_{s=1}^{n}[m_s(x)f(x)]\right)F_p(x)(\text{in } R_p), \\
&= \sum_{s=1}^{n}[pg(x)r_s(x)F_p(x)] + \sum_{s=1}^{n}[m_s(x)f(x)F_p(x)]\,(\text{in } R_p), \\
&= \sum_{s=1}^{n} m_s(x)\,(\text{in } R_p).
\end{aligned} \tag{47}$$

## 3.3 計算實例

本節提供基於晶格後量子密碼學同態加密實例說明；其中，為能對應和方便讀者理解，採用的相關參數值與第 II 節一致，包含如下：$N = 7$、$p = 3$、$q = 128$，值得注意的是這只是計算示意，在真實應用要採用更大的參數值來提升安全性和明文長度。

### 3.3.1 金鑰產製

金鑰值假設與第 II 節一致，$f(x)$的編碼是1||-1||1||0||0||-1||1，$F_p(x)$的編碼是0||2||0||0||1||0||1，$F_q(x)$的編碼是87||58||81||54||36||67||2，並且私鑰的編碼是$f(x)||F_p(x)||F_q(x)$，分別如公式(48)、公式(49)、公式(50)所示。並且公鑰$h(x)$的編碼是12||94||20||56||123||124||83，如公式(51)所示。

$$f(x)(\text{in } R) \equiv x^6 - x^5 + x^2 - x + 1. \tag{48}$$

$$F_p(x)(\text{in } R_p) \equiv x^6 + x^4 + 2x. \tag{49}$$

$$\begin{aligned}&F_q(x)(\text{in } R_q)\\&\equiv 2x^6 + 67x^5 + 36x^4 + 54x^3 + 81x^2 + 58x + 87.\end{aligned} \tag{50}$$

$$\begin{aligned}&h(x)(\text{in } R_q)\\&\equiv 83x^6 + 124x^5 + 123x^4 + 56x^3 + 20x^2 + 94x + 12.\end{aligned} \tag{51}$$

### 3.3.2 加密

本節假設有 2 筆資料，明文編碼分別為1||1||0||0||0||0||0、0||0||1||0||0||0||0，並且解碼為待加密訊息多項式$m_1(x)$和$m_2(x)$ (如公式(52)和公式(53)所示)。其中，待加密訊息多項式$m_1(x)$的加密計算如第 II 節所述，搭配隨機多項式$r_1(x)(\text{in } R) \equiv x^5 + 127x^4 + x^3 + 127$，可得第 1 筆資料的密文$c_1(x)$，如公式(54)所示，並且可以編碼為98||18||58||119||126||82||13。相同的，待加密訊息多項式$m_2(x)$的加密計算如第 II 節所述，搭配隨機多項式$r_2(x)(\text{in } R) \equiv 127x^6 + 127x^5 + x^3 + x^1$，可得第 2 筆資料的密文 $c_2(x)$，如公式(55)所示，並且可以編碼為20||52||123||123||85||16||94。

$$m_1(x)(\text{in } R_p) \equiv x + 1. \tag{52}$$

$$m_2(x)(\text{in } R_p) \equiv x^2. \tag{53}$$

$$\begin{aligned} E(m_1, r_1) &= c_1(x)(\text{in } R_q), \\ &\equiv 13x^6 + 82x^5 + 126x^4 + 119x^3 + 58x^2 + 18x + 98. \end{aligned} \tag{54}$$

$$\begin{aligned} E(m_2, r_2) &= c_2(x)(\text{in } R_q), \\ &\equiv 94x^6 + 16x^5 + 85x^4 + 123x^3 + 123x^2 + 52x + 20. \end{aligned} \tag{55}$$

可以根據公式(45)執行加法同態加密,把公式(54)和公式(55),計算結果$c_1(x) + c_2(x)$如公式(56)所示。

$$\begin{aligned} \sum_{s=1}^{2} E(m_s, r_s)(\text{in } R_q) &= c_1(x) + c_2(x)(\text{in } R_q), \\ &\equiv (13x^6 + 82x^5 + 126x^4 + 119x^3 + 58x^2 + 18x + 98) \\ &\quad + (94x^6 + 16x^5 + 85x^4 + 123x^3 + 123x^2 + 52x + 20) \\ &\qquad (\bmod\ 128)(\bmod\ x^7 - 1), \\ &\equiv 107x^6 + 98x^5 + 83x^4 + 114x^3 + 53x^2 + 70x + 118, \\ &\equiv C(x)(\text{in } R_q). \end{aligned} \tag{56}$$

### 3.3.3 解密

在解密過程中,可以根據公式(46)執行解密計算,把公式(56)的計算結果$c_1(x) + c_2(x)$代入公式(46),可以得到計算結果$m_1(x) + m_2(x)$,如公式(57)所示。通過結果可以觀察到由於隨機多項式$r_1(x)$和$r_2(x)$在加密過程中有乘上$p$,所以在解密過程中將會被消除。因此,這個方法即使是同一把公鑰對相同訊息加密,每次可以得到不同密文,而密文即使加總後仍可消除隨機多項式。

$$D\left(\sum_{s=1}^{n} c_s(x)\right)(\text{in } R_p) = D(C(x))(\text{in } R_p),$$

$$= [C(x) \times f(x)(\text{in } R_q)] \times F_p(x)(\text{in } R_p).$$

$$= \begin{bmatrix} (107x^6 + 98x^5 + 83x^4 + 114x^3 + 53x^2 + 70x + 118) \\ \times (x^6 - x^5 + x^2 - x + 1)(\text{in } R_q) \end{bmatrix}$$

$$\times (x^6 + x^4 + 2x)(\text{in } R_p), \tag{57}$$

$$\equiv (12x^6 - 10x^5 + 13x^4 - 12x^3 + 4x^2 - 2x - 2)(\text{in } R_p)$$

$$\times (x^6 + x^4 + 2x)(\text{in } R_p),$$

$$\equiv (2x^5 + x^4 + x^2 + x + 1)(x^6 + x^4 + 2x)(\text{in } R_p),$$

$$\equiv x^2 + x + 1 = m_1(x) + m_2(x).$$

## 3.4 安全性討論

本節對基於晶格後量子密碼學同態加密的安全性進入深入的討論。

(1). 由於 NTRU 加密演算法屬於一種基於晶格後量子密碼學方法，具備抵抗量子計算攻擊的能力。因此，不論傳統電腦或量子電腦都無法從公鑰破解出其私鑰，也無法從密文破解出明文。

(2). 由於產製公鑰的過程中，私鑰多項式$F_q(x)$將會乘上隨機多項式$g(x)$，所以即使是相同私鑰的情況下，也可以產製不同的公鑰，達到 IND-CPA 或 IND-CCA 安全等級。

(3). 由於加密計算的過程中，公鑰多項式$h(x)$將會乘上隨機多項式$r(x)$再加上明文多項式$m(x)$，所以即使是同一把公鑰對相同明文加密的情況下，也可以產製不同的密文，達到 IND-CPA 或 IND-CCA 安全等級。

(4). 由於每個密文都搭配不同的隨機多項式$r(x)$來產製，從而保障密文的安全性。而在解密過程中會從$R_q$環轉換到$R_p$環，可以把的 $p$ 倍數消除，所以可以消除隨機多項式$r(x)$的影響。

# 4.結論與未來研究

本研究的主要設計基於晶格後量子密碼學同態加密，通過數學理論證明基於晶格後量子密碼學同態加密的可行性，並且通過計算實例演示計算過程和結果，以及提供安全性的證明。可以提供一個具備抵抗量子計算攻擊的同態加密解決方案。

在未來研究，可以考慮把現行的同態加密應用和服務改為後量子密碼學基礎，以提升抵抗量子計算攻擊的能力。並且可以把此研究方法應用到聯邦學習[16]-[17]，以提升資料安全性，並且避免"先竊取，後解密(Harvest Now, Decrypt Later, HNDL)"攻擊。

# 參考文獻